\documentclass[a4paper,12pt,showpacs,preprintnumbers,amsmath,amssymb]{revtex4}
\usepackage{amsmath}
\usepackage[dvips]{graphicx,color}

\usepackage{amssymb}
\usepackage{array,tabularx}
\usepackage{indentfirst}

\begin{document}

\tolerance=5000

\def\pp{{\, \mid \hskip -1.5mm =}}
\def\cL{{\cal L}}
\def\be{\begin{equation}}
\def\ee{\end{equation}}
\def\bea{\begin{eqnarray}}
\def\eea{\end{eqnarray}}
\def\beq{\begin{eqnarray}}
\def\eeq{\end{eqnarray}}
\def\tr{{\rm tr}\, }
\def\nn{\nonumber \\}
\def\e{{\rm e}}

\def\ben{\begin{enumerate}}
\def\een{\end{enumerate}}
\def\bei{\begin{itemize}}
\def\eei{\end{itemize}}
\def\ni{\noindent}
\def\bs{\bigskip}
\def\ms{\medskip}


\title{Initial and final de Sitter universes from modified $f(R)$ gravity}

\author{G. Cognola$^{1}$, E. Elizalde$^{2}$, S.D. Odintsov$^{3}$,
P. Tretyakov$^{4}$ and S. Zerbini$^{1}$}

\affiliation{$^{1}$Dipartimento di Fisica, Universit\`a di Trento
and Istituto Nazionale di Fisica Nucleare
Gruppo Collegato di Trento, Italia}

\affiliation{$^{2}$Consejo Superior de Investigaciones Cient\'\i
ficas ICE/CSIC-IEEC, Campus UAB, Facultat de Ci\`encies, Torre
C5-Parell-2a pl, E-08193 Bellaterra (Barcelona) Spain}

\affiliation{$^{3}$Instituci\`{o} Catalana de Recerca i Estudis Avan\c{c}ats
(ICREA) and Institut de Ciencies de l'Espai (IEEC-CSIC), Campus UAB,
Facultat de Ci\`encies, Torre C5-Parell-2a pl, E-08193 Bellaterra
(Barcelona) Spain}

\affiliation{$^{4}$JINR, Dubna, Moscow region, Russia}

\date{{\small \today}}


\begin{abstract}
Viable models of modified gravity which satisfy both local as well
as cosmological tests are investigated. It is demonstrated that
some versions of such highly non-linear models exhibit multiply de Sitter
universe solutions, which often appear in pairs, being one of them stable
and the other unstable. It is explicitly shown that, for some values of the
parameters, it is possible to find several de Sitter spaces
 (as a rule, numerically); one of them may serve for the inflationary stage,
 while the other can be used for the description of the dark energy epoch.
 The numerical evolution of the effective equation of state parameter
 is also presented, showing that these models can be considered
 as natural candidates for the unification of early-time inflation
 with late-time acceleration through dS critical points. Moreover,
based on the de Sitter solutions, multiply SdS universes are constructed
which might also appear at the (pre-)inflationary stage.
Their thermodynamics are studied and free energies are compared.
\end{abstract}

\pacs{11.25.-w, 95.36.+x, 98.80.-k}

\maketitle

\section{Introduction}

Modified gravity (for a review, see e.g.~\cite{review}) constitutes an interesting dynamical
alternative to $\Lambda$CDM cosmology in that it is also able to describe with success the current
acceleration in the expansion of our Universe, the present dark energy epoch. The specific class of
modified $f(R)$ gravities (for a review, see e.g.~\cite{review,review1}) has undergone many studies
which suggest that this family of gravitational alternatives for dark energy is able to pass the
constringent solar system tests. The investigation of cosmic acceleration
 as well as the study of the cosmological properties of $f(R)$ models has
been done in Refs.~\cite{review,review1,NO,FR,FR1,cap}. The possibility of a
natural unification of early-time inflation with late-time acceleration becomes a
realistic and quite natural possibility in such models, as is demonstrated
e.g. in Ref.~\cite{NO}.

Recently, the importance of modified gravity models of this kind has been reassessed with the
appearance of the so-called `viable' $f(R)$ models \cite{HS,AB,Uf,UUprd,Cognola:2007zu}. Those are
theories which satisfy both the cosmological as well as the local gravity constraints, which had
caused in the past a number of serious problems to some of the first-generation theories, that had
to be considered now as only approximate descriptions from more realistic theories. The final aim of
all those phenomenological models is to describe a segment as large as possible of the entire
history of our universe, as well as to recover all local predictions of Einstein's gravity that have
been already verified experimentally, with very good accuracy, at the solar system scale. It is
remarkable that, as was demonstrated in Refs.~\cite{Uf,UUprd}, some of these realistic models lead
to a natural unification of the early-time inflation epoch with the late-time acceleration stage.

Let us recall that, in general (see e.g. \cite{review,review1}, for a
review), the total
action for the modified $f(R)$ gravitational models can be written as
\be
\label{XXX7}
S=\frac{1}{\kappa^2}\int d^4 x \sqrt{-g} \left[R-F(R)\right]+S_{(m)}\, .
\ee
Here $F(R)$ is a suitable function of the scalar curvature $R$, 
which defines the modified gravitational
part of the model. The general equation of motion in $f(R)\equiv R-F(R)$
gravity with matter is given by
\be
\label{XXX22}
\frac{1}{2}g_{\mu\nu} f(R) - R_{\mu\nu} f'(R) - g_{\mu\nu} \Box f'(R) +
\nabla_\mu \nabla_\nu f'(R)
= - \frac{\kappa^2}{2}T_{(m)\mu\nu}\ ,
\ee
where $T_{(m)\mu\nu}$ is the matter energy-momentum tensor and 
$f'(R)$ is the derivative of $f(R)$ with respect to its argument $R$.
For a generic $f(R)$ model is not easy to find exact static solutions.
However, if one impose some restrictions, one can proceed along the following
lines. First, we may require the existence of solutions with
{\it constant} scalar curvature $R=R_0$, and we arrive at
\be
f'(R_0)R_{\mu\nu}=\frac{f(R_0)}{2}g_{\mu\nu}\,.
\label{geneqRc}
\ee
Taking the trace, we have the condition
\be
2f(R_0)=R_0\,f'(R_0)
\label{B}
\ee
and this means that the solutions
are Einstein's spaces, namely they have to satisfy the equation
\be
R_{\mu\nu}=\frac{f(R_0)}{2 f'(R_0)}g_{\mu\nu}=
\frac{R_0}{4}g_{\mu\nu}\,,
\label{E}
\ee
 $R_0$ being a solution of
Eq.~(\ref{B}). This gives rise to an effective cosmological constant, namely \be
\Lambda_{eff}=\frac{f(R_0)}{2 f'(R_0)}=\frac{R_0}{4}\,. \label{l} \ee The purpose of our work will
be to study the appearance of multiply de Sitter space solutions in several realistic models of
modified gravity. The occurrence of multiply de Sitter solutions plays a fundamental role in
modified gravity because it permits to describe the inflation stage as well as current $\Lambda$CDM
cosmology in terms of this theory alone without any need for either fine-tuning of a cosmological
constant nor of introducing extra scalar fields. In summary, this is a minimal and at the same time
very powerful approach which circumvents some of the hardest problems of present day physics. One
may argue that these theories are equivalent to introducing extra scalar fields, but this
equivalence has been proven to hold at the classical level only, not at the quantum one. In
addition, the cosmological interpretation of modified gravity solutions is different from that of
scalar field cosmology.

The paper is organized as follows. In the next section we discuss a viable modified gravity model
\cite{HS} in an attempt to study de Sitter solutions there. It is shown that, for some values of the
parameters, it is possible to find several de Sitter spaces (as a rule, numerically, for small
values of the curvature exponent); one of them may serve for the inflationary stage, while the other
one can be used for the description of the dark energy stage. The evolution of the effective
equation of state parameter is investigated numerically. Sect.~3 is devoted to the analysis of the
same question in a slightly generalized model which is known to describe the unification of the
early-time inflation with the late-time acceleration epochs \cite{Uf}. The same numerical
investigation is carried out, and a multiply de Sitter universe solution is constructed. 
In section 4, the corresponding problem is investigated 
for a viable model of tangential modified gravity,
proposed in Ref.~\cite{Cognola:2007zu}. Having in mind the possibility to use a SdS universe for the
description of cosmic acceleration, those solutions for modified gravity are investigated in
Sect.~5. Comparison of the free energies of the de Sitter and the SdS solutions, for the viable
model of the second section, is then made. Finally, some outlook is given, together with the
conclusions, at the end of the paper.

\section{Dynamical system approach and de Sitter solutions in
realistic modified gravity}

In this section we will study de Sitter solutions in realistic
modified gravity models using a dynamical system approach.
We shall start from the following form for the initial action
\begin{equation}
S=\int d^4 x \sqrt{-\mathrm{g}} \left [ \frac{1}{2\kappa^2}R
 -F(A)\right ],
 \label{1.1}
\end{equation}
where $A$ is some function of the geometrical invariants.
In order to make use of the method of dynamical systems, we work in the
metric corresponding to a spatially-flat FRW universe, namely
\begin{equation}
\mathrm{g}_{\mu\nu}=diag(-n(t)^2, a(t)^2 , a(t)^2 , a(t)^2).
 \label{1.2}
\end{equation} 
From here, the FRW equation can be written in the following
way (for more details see Ref.~\cite{NOT})
\begin{equation}
\frac{3}{\kappa^2}H^2=\rho_{F},
 \label{1.3}
\end{equation}
 where
\begin{equation}
\rho_{F}= F +F'A_n - 3HF'A_{\dot n} - F'\frac{d A_{\dot n}}{dt} -F''A_{\dot n}\dot A.
 \label{1.4}
\end{equation}
The ``dot'' over the symbol means derivative with respect to time $t$, while
$A_n$ and $A_{\dot n}$ represent the derivative of $A$ with respect to
$n$ and $\dot n$ respectively. 
In what follows we will mainly concentrate on the simplest case $A=R$.

Let us consider the following choice for the function $F$ \cite{HS},
which represents a very interesting subclass of viable modified gravitational
models \begin{equation}
F(R)=\frac{\mu^2}{2\kappa^2}\frac{c_1(\frac{R}{\mu^2})^k+c_3}{c_2(\frac{R}{\mu^2})^k+1},
 \label{1.5}
\end{equation}
where the constant $\mu$ has dimension of mass, while 
$c_1$, $c_2$, $c_3$ are some positive dimensionless constants.
Note that here $R=\frac{6}{n^2}\left[
\frac{\ddot a}{a} - \frac{\dot a}{a}\frac{\dot n}{n}+\frac{\dot
a^2}{a^2} \right ]$, $R_n=-2R=-12(\dot H+2H^2)$, $R_{\dot n}=-6H$.
After making variation, it is chosen $n=1$.
Introducing a nonzero constant $c_3$ is here equivalent to
introducing a shift in the effective cosmological constant,
 and for this reason we will assume
$c_3=0$ in our further computations. Finally, we can write Eq.~(\ref{1.3})
for the function in (\ref{1.5}) as
\begin{equation}
\begin{array}{l}
\frac{6}{\mu^2}H^2= \frac{c_1\left(\frac{R}{\mu^2}\right)^k}{\left(c_2\left(\frac{R}{\mu^2}\right)^k+1\right)} - \frac{6(\dot H+H^2)\frac{kc_1}{\mu^2}\left(\frac{R}{\mu^2}\right)^{k-1}}{\left(c_2\left(\frac{R}{\mu^2}\right)^k+1\right)^2}+\\
\\
 \frac{36(H\ddot H+4H^2\dot H)}{\left(c_2\left(\frac{R}{\mu^2}\right)^k+1\right)^3}\frac{kc_1}{\mu^4}\left [(k-1)\left (\frac{R}{\mu^2}\right )^{k-2}-c_2(k+1)\left(\frac{R}{\mu^2}\right)^{2k-2} \right ],
 \end{array}
 \label{1.6}
\end{equation}
where $R=6(\dot H+2 H^2)$. This equation can be rewritten as a dynamical system, namely
\begin{equation}
\begin{array}{l}
\dot H =C,\\
\dot C =F_1(H,C).
\end{array}
 \label{1.7}
\end{equation}
 It is easy to see that the critical points of this system are the de
Sitter points ($\dot H=0,\,\ddot H=0$). To investigate the nature of
these points we need to determine
 them explicitly, a non-trivial problem in the general case.
It is however easy to obtain from (\ref{1.6}) the equation satisfied by the
critical points $H_0$:
\begin{equation}
\frac{6}{\mu^2}H_0^2= \frac{c_1\left(\frac{12H_0^2}{\mu^2}\right)^k}{\left(c_2\left(\frac{12H_0^2}{\mu^2}\right)^k+1\right)} - \frac{6H_0^2\frac{kc_1}{\mu^2}\left(\frac{12H_0^2}{\mu^2}\right)^{k-1}}{\left(c_2\left(\frac{12H_0^2}{\mu^2}\right)^k+1\right)^2}.
 \label{1.8}
\end{equation}
The same result can be obtained directly starting from equation (\ref{B})
of the previous Section.

It is convenient to introduce the notation $x_0\equiv 12\frac{H_0^2}{\mu^2}$, for further investigation of this equation. So, finally, we have
\begin{equation}
c_2^2x_0^{2k+1}-2c_1c_2x_0^{2k}+2c_2x_0^{k+1}+2c_1(\frac{1}{2}k-1)x_0^k+x_0=0.
 \label{1.9}
\end{equation}
First of all we find one (trivial) root of this equation $x_0=0$ (it corresponds to $H_0=0$), what allows us to reduce the order of the equation.
(Note also that, if $c_3\neq 0$, Eq.~(\ref{1.9}) takes a more complicate
form and then it does not have the trivial root $x_0=0$.) But nevertheless the equation still is of
 $2k$-order and in the interesting case ($k>2$); this is too high and the roots cannot be found algebraically. We need a specific discussion of this problem. It is clear that
 if there are too many (say 10 or 20) de Sitter points in the theory, it
looks like a classical analogue (at a reduced scale, of course) \cite{class}
 of the string landscape vacuum structure in which case it will be far from trivial to
obtain the standard cosmology. Nevertheless, something can be
done even in this case, by comparing the energies of the corresponding
de Sitter solutions, which should in fact differ, as is discussed in Sect.~V.
 Using Descartes rule of signs, we find that Eq.~(\ref{1.9}) can have 2 or 0 positive roots.
 More detailed information can be obtained by using Sturm's theorem. Unfortunately it is
 not possible to apply it in the general case (for arbitrary $k$), so we need to
 investigate separately each of the different
cases, $3\leqslant k\leqslant 10$, and see which one is more interesting
from a cosmological viewpoint. We have therefore undertaken here a systematic
analysis of all possible cases in this region of values of $k$.

 Computations are rather involved (specially for larger $k$), but the final
 results are not so difficult to describe analytically. We have found, for any of the values of $k$ considered, in the range above, that the number of roots depends on the parameter $\alpha_k=\frac{c_1^k}{c_2^{k-1}}$ only,
 and that there is an $\alpha_k^*$ such that, for $0<\alpha_k<\alpha_k^*$, Eq.~(\ref{1.9}) has no positive roots, while for $\alpha_k>\alpha_k^*$, Eq.~(\ref{1.9}) has two positive roots.
We here now enumerate our results systematically: $\alpha_3^*=-\frac{3^3 11}{2}+\frac{3^3 5}{2}\sqrt{5}\approx2.43$;
 $\alpha_4^*=\frac{1237}{2^5}-\frac{1837}{2^5 3^2}\sqrt{33}\approx2.01$; $\alpha_5^*=-\frac{5^5}{2^2 3^4}+\frac{5^5}{3^5 2}\sqrt{3}\approx1.49$;
 $\alpha_6^*=\frac{3^7 17\cdot 6977}{2^{24} 5}-\frac{2\cdot 8329\cdot 32183}{2^{24}5^3}\sqrt{65}\approx1.03$;
 $\alpha_7^*=-\frac{7^7 17681}{2\cdot 3^3 5^{13}}+\frac{7^9 107}{2\cdot 3^2 5^{13}}\sqrt{21}\approx0.68$;
 $\alpha_8^*=\frac{69401}{2\cdot 3^7 7^2}+\frac{5\cdot 66821}{2\cdot 3^8 7^4}\sqrt{105}\approx0.43$;
 $\alpha_9^*=\frac{3^{18}\cdot 719\cdot 24709}{2^{8} 7^{17}}+\frac{3^{18} 517399}{2^{3} 7^{17}}\sqrt{2}\approx0.26$;
 $\alpha_{10}^*=\frac{5^{10}401843202307}{2^{59}3^4}+\frac{5^{10} 17\cdot 67\cdot 179\cdot 659\cdot 164429}{2^{59}3^9}\sqrt{17}\approx0.16$.
 Note also that, near the critical point $\alpha_k^*$, both solution are very close to each other when they exist (a real value of $\alpha_k>\alpha_k^*$),
 and they are complex conjugate to each other for $0<\alpha_k<\alpha_k^*$. This means that, in a situation where these two roots differ substantially, it must necessarily be $\alpha_k\gg\alpha_k^*$.

The dS-points described above can be used for the construction of inflationary
or late-time acceleration behavior (depending on the value of the scale
factor $\mu^2$). To this aim, the corresponding dS point must be unstable in the
inflationary stage but can be either unstable
 or stable for the late-time acceleration one.
However, this is in fact not a strictly necessary condition, since even for stable dS
inflation, the exit from it can be achieved by a coupling with matter,
through the effect of small non-local term or by some other mechanism.
Unfortunately, an all-round investigation of stability
of the dS point turns out to be very difficult in the general case.
For the recent analysis of critical points in more general 
modified gravity theories depending on all geometrical invariant  
see Ref.~\cite{Cognola:2008wy}.

\begin{figure}[ht]

\includegraphics[width=0.5\textwidth]{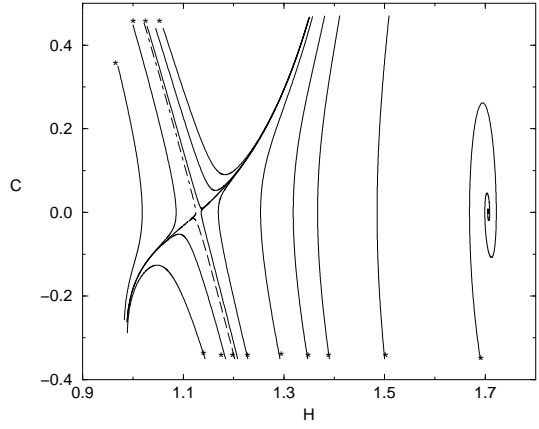}

 \caption{Phase plane ($H-C\equiv\dot H$) near the equilibrium points. Stars denote the
 beginning of the trajectories.
 Dashed and dot dashed lines lead to bottom.}
 \label{Fig.1}

 \end{figure}

Here we can carry out the analysis on the stability of de Sitter points
for the model we are dealing with in this section, only
for some chosen numerical set of parameters. Such computations show that
 one of the dS points is very likely to be always unstable (with a smaller
value of $H$) and that another one is very likely to be always stable
(with
larger value of $H$). This means that there is actually an easy way for the
inflationary stage construction (if the initial conditions lie
sufficiently close
to the unstable dS point, see Fig.~\ref{Fig.1}). However, only within the
model under consideration it might be a
problem with late-time acceleration, originated from the stable point. The only
possibility for late time acceleration is to choose very special parameters,
in which the stable point is situated very far away from the unstable one.
Of course, the problem disappears when one describes only late-time
acceleration within such a model, or when one takes
into account other terms like the local ones. Note however that specific values
of the parameters in which both points are unstable can be chosen too,
 which give rise to a sound theory which provides a unified description of
the inflationary and late-time acceleration epochs.

 \begin{figure}[ht]

\includegraphics[width=0.5\textwidth]{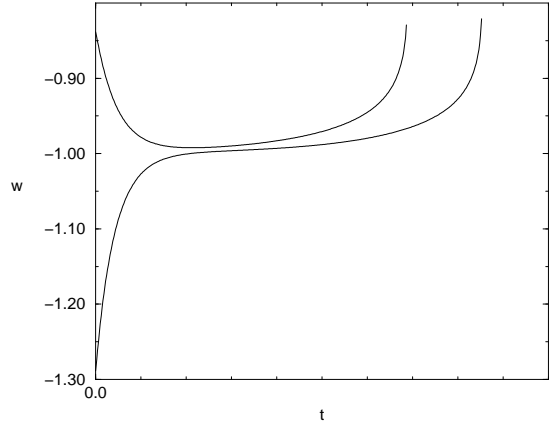}

 \caption{Evolution of $w_{eff}$ near an unstable point for the dashed and
 dot-dashed lines of Fig.~\ref{Fig.1}. The time origin corresponds to the stars from the previous graph.}
 \label{Fig.2}

 \end{figure}

Let us now consider specific numerical results. First, note that a full numerical investigation starting
from the inflation (or from our dS point) and extending to the late-time acceleration epoch and going
 through a FRW-like stage is not feasible because of the presence of numerical instabilities that prevent
 such possibility. For this reason, we will investigate the regions near the equilibrium points only,
 and especially near the unstable one. In Fig.~\ref{Fig.1} the trajectories in the phase plane ($H-\dot H$)
are presented for the set of parameters: $\mu=1$, $k=4$, $c_1=1.12\cdot 10^{-3}$, $c_2=6.21\cdot 10^{-5}$.
Such small values of $c_1$ and $c_2$ are motivated in \cite{HS}, but the
picture is typical in any case. The position of the dS point is specified. We may see one stable point
 (with bigger value of $H$) and another, unstable one. The stars in the graph correspond to the beginning
 of the trajectories. We see clearly that trajectories can pass sufficiently close to the dS point,
and that they escape from it, lead by the two attractors. One of them (the upper one) leads to a singularity, while another one (heading towards the bottom), leads to a solution going
somewhere in the FRW-like region (this is difficult to analyze exactly, due
to numerical instabilities). In Fig.~\ref{Fig.2} the evolution of
$w_{eff}\equiv -1-\frac{2\dot H}{3H^2}$ is depicted for two trajectories: the upper one corresponds
to the dashed line of Fig.~\ref{Fig.1}, the bottom one to the dot-dashed line of Fig.~\ref{Fig.1}
 (both of them lead to the bottom in Fig.~\ref{Fig.1}). We see that there actually are trajectories
 which start from phantom-like solutions and tend to a normal evolution
(when this is possible in the model). Of course, these are qualitative
considerations only, owing to the fact that no matter is taken into account.
Most probably, inclusion of matter will change the situation. However, it is just
remarkable that both the early-time inflation epoch as well as the late-time
accelerating one could be obtained in a unified and natural way in
such a model, owing to the presence of several dS points.

\section{A different example of $f(R)$ viable model}

A slightly different choice for the function $F(R)$ is motivated by the
realistic and viable model which was proposed for the unification of
early-time inflation and late-time acceleration in Ref.~\cite{Uf},
namely \begin{equation}
F(R)=\frac{(R-R_0)^{m}+R_0^{m}}{f_0+f_1\left[
(R-R_0)^{m}+R_0^{m} \right]}.
 \label{2.1}
\end{equation}
Here we imply that $f_0$, $f_1$ and $R_0$ is positive, what seems reasonable.
Now, our FRW-like equation (\ref{1.3}) reads
\begin{equation}
\begin{array}{l}
\frac{3}{\kappa^2}H^2= \frac{\{(R-R_0)^m+R_0^m\}}{\left(f_0+f_1\{(R-R_0)^m+R_0^m\}\right)} - \frac{6(\dot H+H^2)f_0m(R-R_0)^{m-1}}{\left(f_0+f_1\{(R-R_0)^m+R_0^m\}\right)^2}+\\
\\
 \frac{36(H\ddot H+4H^2\dot H)f_0m(R-R_0)^{m-2}}{\left(f_0+f_1\{(R-R_0)^m+R_0^m\}\right)^3}\left [ (m-1)(f_0+f_1R_0^m)-(m+1)f_1(R-R_0)^m \right ],
 \end{array}
 \label{2.2}
\end{equation}
where $R=6(\dot H+2 H^2)$. Rewriting this equation under the form of a dynamical system (\ref{1.7}) we find, by introducing the new
 variable $x\equiv 12H_0^2-R_0$, that the equilibrium points of this system (which are dS points indeed) are obtained from the following equation
\begin{equation}
\begin{array}{l}
f_1^2x^{2m+1}+f_1(f_1R_0-4\kappa^2)x^{2m}+2f_1(f_0+f_1R_0^m)x^{m+1}+\\
\\ \left[2R_0f_1(f_0+f_1R_0^m)-8\kappa^2f_1R_0^m+2\kappa^2(m-2)f_0\right]x^m+2\kappa^2R_0f_0mx^{m-1}+\\
\\ (f_0+f_1R_0^m)^2x+R_0(f_0+f_1R_0^m) (f_0+f_1R_0^m-4\kappa^2R_0^{m-1})=0.
 \end{array}
 \label{2.3}
\end{equation}
It is clear that, in the general case (even for fixed $m>2$), solving this
equation will not be an easy thing. We can make use of the following hint. Let us require that $x=0$ be a solution of Eq.~(\ref{2.3}). This is possible only if there is some relation among the constants of our theory. As we can easily see from (\ref{2.3}), it must be that $f_0=R_0^{m-1}(4\kappa^2-f_1R_0)$. Moreover, since $f_0>0$, we need that $f_1R_0<4\kappa^2$. Substituting this condition into (\ref{2.3}) we find an essential simplification of the former equation, namely
\begin{equation}
\!\begin{array}{l}
f_1^2x^{2m}-f_1(4\kappa^2-f_1R_0)x^{2m-1}+8\kappa^2f_1R_0^{m-1}x^{m}+\\
\\ 2\kappa^2(m-2)R_0^{m-1}(4\kappa^2-f_1R_0)x^{m-1}+2\kappa^2R_0^m(4\kappa^2-f_1R_0)mx^{m-2}+ (4\kappa^2R_0^{m-1})^2=0,
 \end{array}
 \label{2.4}
\end{equation}
 where the root $x=0$ is already excluded. We now discuss the physical meaning of finding a solution $x=0$. As a rule this will imply that
 $R_0$ is the value of the scalar curvature at present time, but in principle $R_0$ is just a parameter of the theory, which could be given any likely value. We can, for instance, fix $R_0$ to have the dS point value which is situated exactly at present time, $t_0$, or at $t_0+100$ years. And there is also the possibility
 to explain late time acceleration if this point is stable. Now let us consider other possible roots of Eq.~(\ref{2.4}).
 Using Sturm's theorem as in the previous section, we find several positive roots (which correspond to dS points in the past) and a number of negative roots (needless to say, only values which are $x>-R_0$ have a physical meaning), corresponding to dS points in the future.
 Unfortunately, computations are harder than in the previous case and our calculations have been performed for three specific values of the parameter $m$: 3, 5 and 7, only.
 In any case, the results are very similar for the three situations, what gives us hope that for larger values of $m$ result will be also similar. We may
 predict the upper limit of the dS points in the future only, which is 2, but in any case this point is not much interesting.

 Concerning the dS points in the past, the picture turns out to be much like the one in the previous case: the number of dS points depends only on a dimensionless
 parameter for all investigated values of $m$. That is, $\beta = \frac{2\kappa^2}{f_1R_0}$. And there are two values, $\beta_*$ and $\beta_{**}$, which are different
 for each $m$, so for $0<\beta<\beta_*$ and $\beta>\beta_{**}$ there are two dS points and for $\beta_*<\beta<\beta_{**}$ there is no dS point. We have not been able to obtain exact analytical expressions for $\beta_*$ and $\beta_{**}$ and thus give here numerical results only. For $m=3$
 $\beta_*\approx 0.153$, $\beta_{**}\approx 3.12$ (as positive roots of $4\beta^4+336\beta^3-843\beta^2-815\beta+144=0$).
 For $m=5$ $\beta_*\approx 0.163$, $\beta_{**}\approx 2.01$ (as positive roots of
 $1318032\beta^8+40509072\beta^7-208593144\beta^6+472402800\beta^5-637548615\beta^4 +422702939\beta^3-90089631\beta^2-69190983\beta+12301875=0$).
 For $m=7$ $\beta_*\approx 0.215$, $\beta_{**}\approx 1.68$ (as positive roots of
 $2^6 5^9 24337\beta^{11}+2^6 5\cdot181772596417 \beta^{10}-2^4 7 \cdot17\cdot29\cdot641\cdot15074567\beta^9+
 2^5 19\cdot29\cdot173\cdot6421\cdot100207\beta^8-2^2 7\cdot153140941633579\beta^7+2^2 7\cdot227057313819467\beta^6-
 5\cdot7\cdot195531002131489\beta^5+2^2 47\cdot1331611\cdot21800759\beta^4-2\cdot5\cdot53\cdot2797\cdot2102824303\beta^3+
 2^3 7\cdot29\cdot728096696819\beta^2-5^4 23\cdot18382295597\beta+2^{16} 5^5 7^6=0$).

 As we can see, the interval where there are no dS points shrinks when
 the value of the parameter $m$ is increased. If we assume that
 for larger $m$ the number of dS points depends on the parameter $\beta$ only---as clearly happens in the investigated cases---we do find
 $\beta_*$ and $\beta_{**}$ for any value of $m$. Such investigations show that $\beta_*$ and $\beta_{**}$ slowly change when $m$
 increases and that the interval without dS points is present even for very big value of the parameter $m$. For example, for $m=101$ we have $\beta_{*}\approx0.480$ and $\beta_{**}\approx1.06$.
 A simple numerical investigation of stability for the existing
 dS points shows that one of them (the one with the smaller value of $H$) is always unstable, but the other one (with the larger value of $H$) can either be stable or unstable. From our numerical results, the appearance of unstable dS points pairs relative to the interval $0<\beta<\beta_*$ seems most probable. Note also that the condition $f_0>0$ means that it actually must be $\beta>0.5$ and that in this case it is most likely the one stable and one unstable point show up.

 Now let us consider the evolution equation (\ref{2.2}) from a different point of view. As we already noted, $R_0$ is a parameter of the theory which corresponds to the value of the scalar curvature at some epoch. This means that during the normal
 evolution of our universe, from large to little (or zero) curvature, it becomes $R=R_0$ at some moment. But, as we can see from Eq.~(\ref{2.2}), this means that that at this point the coefficient
of the higher derivative term ($\ddot H$) is equal to zero.
 This is a well-known mathematical problem, which needs special
investigation. As we know from mathematics there are two possibilities: the solution of the perturbed equation (with a higher derivative term) may tend to the solution of the
degenerate equation (without higher derivatives), and
 then the coefficient of the higher derivative term may tend to zero or
not tend to the solution of the degenerate equation. A special
 investigation of this problem shows that solutions of Eq.~(\ref{2.2})
 tend to solutions of the degenerate equation, which is
 \begin{equation}
\frac{3}{\kappa^2}H^2_d= \frac{R_0^m}{\left(f_0+f_1R_0^m\right)},
 \label{2.5}
\end{equation}
when $R=R_0+0$ but not so when $R=R_0-0$. This means that the point $R_0$ is reachable during the evolution from large $R$ to zero, but it is not the final point of the evolution, because there is an instability in the future.
On the other hand, we have $R=R_0$ at this point, and therefore $3H_d^2=R_0/4$. Substituting this into (\ref{2.5}) we find a relation
among the parameters of the theory which were introduced before, $f_0=R_0^{m-1}(4\kappa^2-f_1R_0)$. This means that, generically, if one wants the point $R_0$ to be reachable during evolution, a relation of this sort among the parameters of the theory must be fulfilled. That is, we
have two independent parameters only (or even just one if we consider that $R_0$ is strictly related to our epoch). Note also that it is impossible to use standard numerical methods near this point for solving Eq.~(\ref{2.2}) because all available methods need solving the equation with respect to the highest derivative, and it turns out that, near this point, the numerical solution is unstable.

\begin{figure}[ht]

\includegraphics[width=0.5\textwidth]{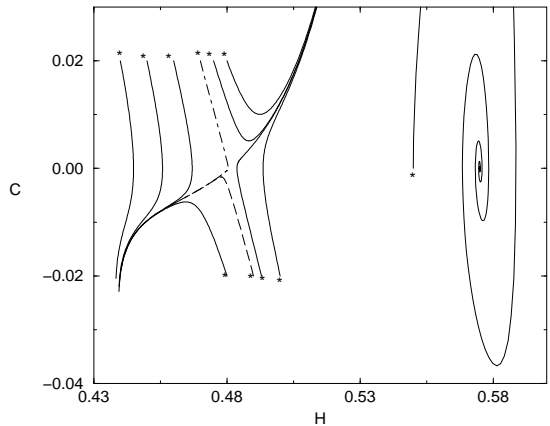}

 \caption{Phase plane ($H-C\equiv\dot H$) near equilibrium points. Stars denote beginning
 of trajectories.
 Dashed and dot dashed lines lead to bottom.}
 \label{Fig.3}

 \end{figure}

 \begin{figure}[ht]

\includegraphics[width=0.5\textwidth]{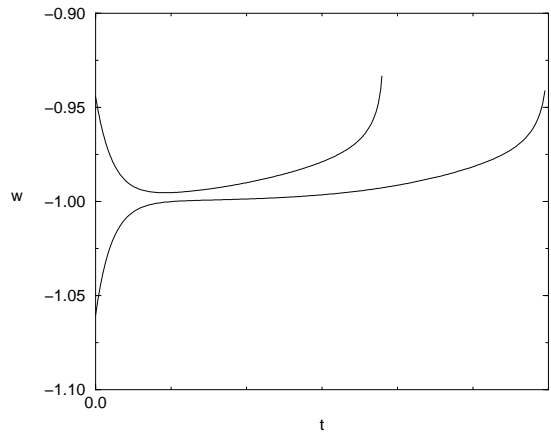}

 \caption{Evolution of $w_{eff}$ near unstable point for dashed and dot-dashed
 lines of Fig.\ref{Fig.1}. Zero in time corresponds to the stars from the previous graph.}
 \label{Fig.4}

 \end{figure}

 Now let us consider numerical results near the stable points. Since the
value of $R_0$ is very small in our epoch ($\sim 10^{-56}$ cm$^{-2}$), it is impossible to use real values in computations. Thus, we have produced a qualitative analysis using the following set of parameters: $R_0=1$, $f_1=1$, $\kappa^2=1$, $m=7$. As we can see, $\beta =2$, and there must be two dS points corresponding to positive $x$. The evolution lines near these points are represented in Fig.~\ref{Fig.3}, where we can trace one stable and one unstable point, stars denoting the beginning of trajectories, as in the previous case. Actually, there are three dS points for these chosen parameters. One of them, $H=0.288$, which corresponds to $x=0$, is degenerate and situated out of the graph. The time evolution of $w_{eff}$ is represented in Fig.~\ref{Fig.4} by the dashed and dot-dashed lines. The picture is very similar to the one in the previous section, and the same comments also apply here.

Thus, we have here shown that the unification of early-time inflation with late-time acceleration is
in principle possible, and even very likely, due to the appearance of several dS points in the
evolution of the universe. For a more realistic study, the presence of (minimally and non-minimally
coupled) matter should be taken into account with care.

\section{de Sitter universe from tangential modified gravity}

\begin{figure}[ht]
\includegraphics[width=0.4\textwidth]{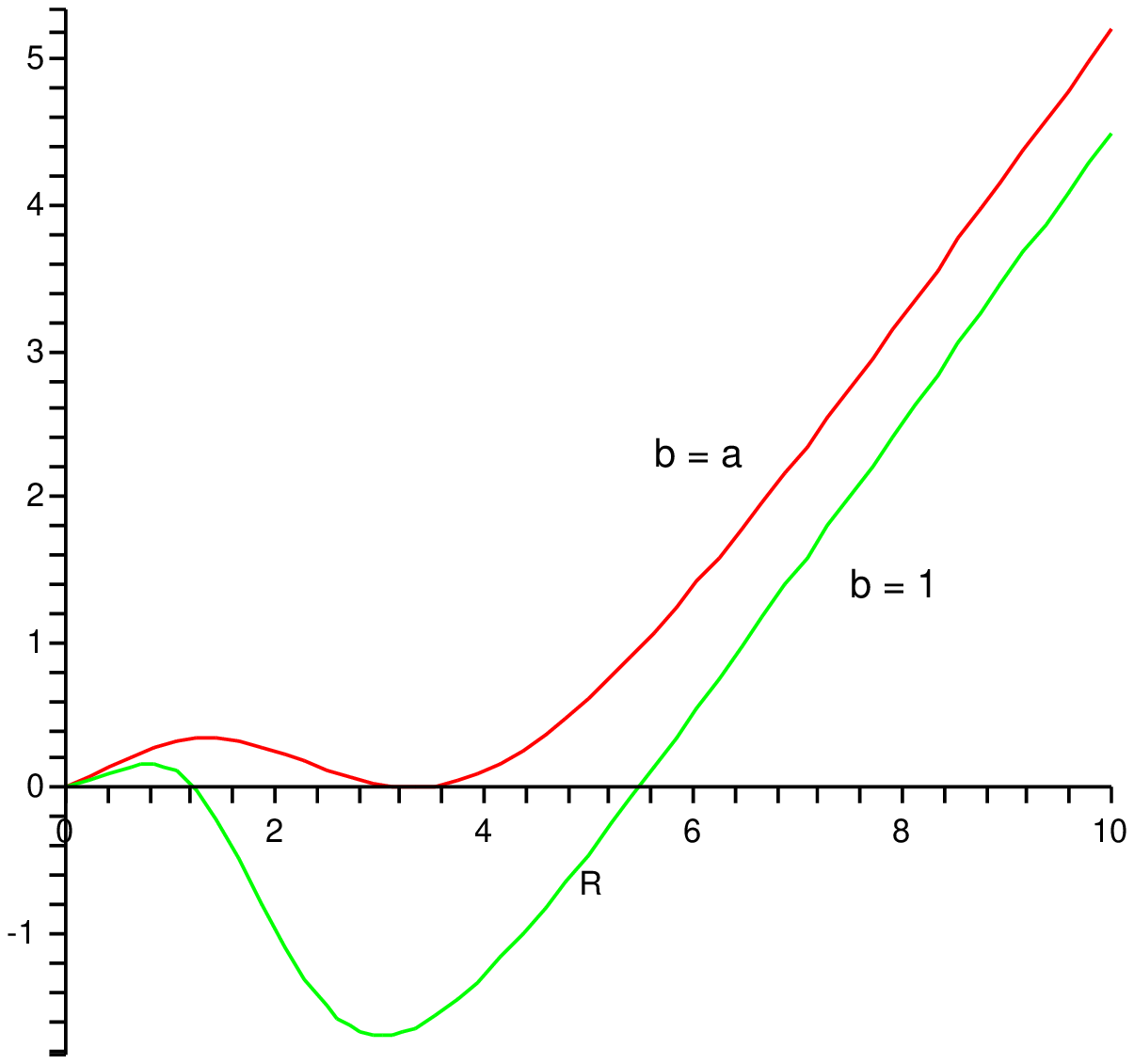}
\includegraphics[width=0.4\textwidth]{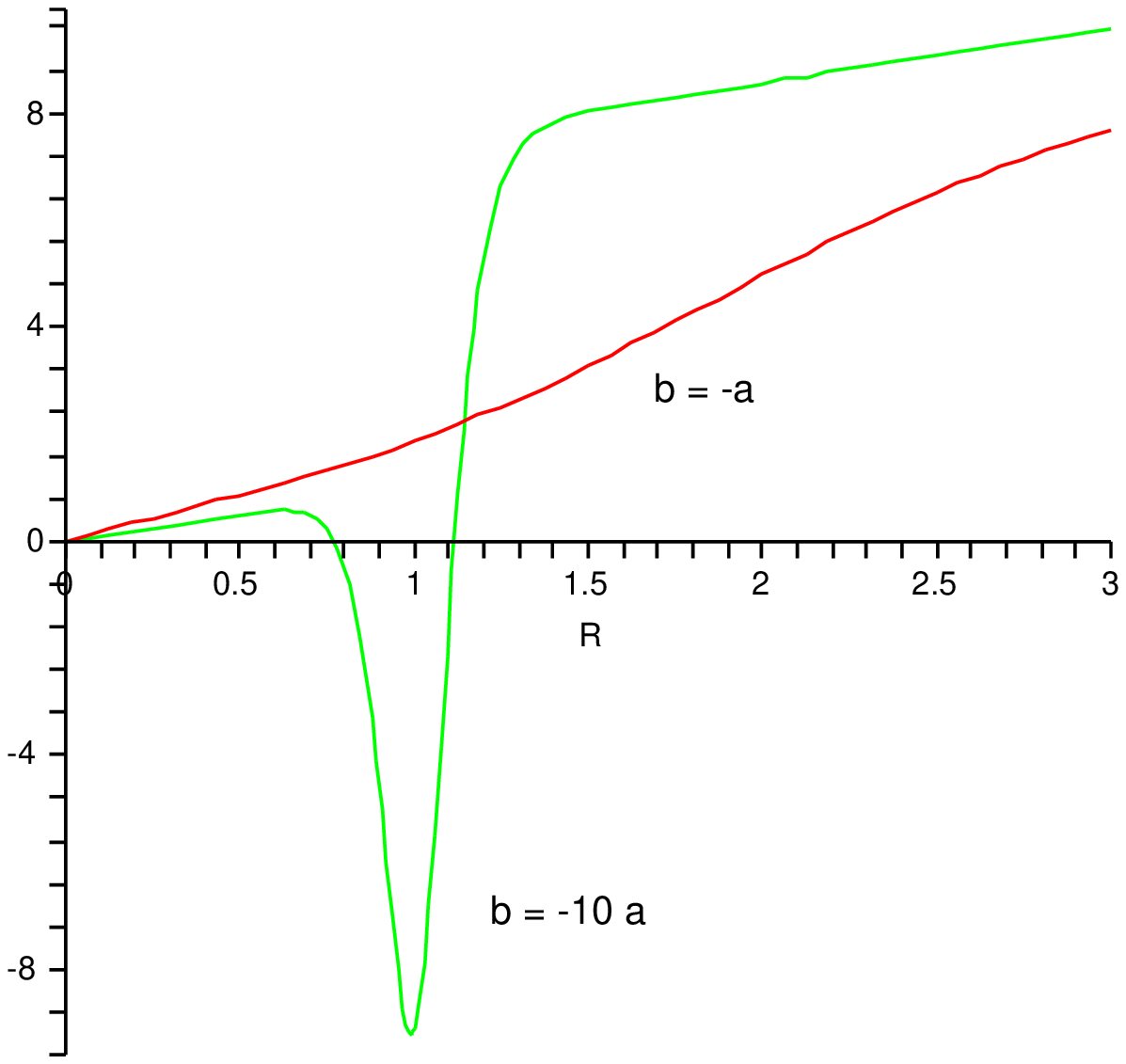}
 \caption{$K(R)$ for $R_0=1,a=1.64$
 and different choices of the parameter $b$:
 $b=1$, $b=a$ (picture on the left);
 $b=-a$, $b=-10a$ (picture on the right).}
 \label{P:tanh1}
 \end{figure}

As a third example, now we discuss a model proposed in \cite{Cognola:2007zu}.
Such model is defined by means of the function
\beq
\label{tan1}
f(R)=R-F(R)=R-a\left[\tanh\left(\frac{b\left(R-R_0\right)}{2}\right)
 +\tanh\left(\frac{b R_0}{2}\right)\right]\,,
\eeq
where $a$, $b$ and $R_0$ are arbitrary parameters.
One immediately sees that $F(0)=0$, as required and, moreover, that
\beq
\label{tan3}
\lim_{R\to\infty}F(R)=2\Lambda_{\rm eff}
 \equiv a\left[1+\tanh\left(\frac{b R_0}{2}\right)\right]\,.
\eeq
If $R\gg R_0$ in the present universe then $\Lambda_{\rm eff}$
plays the role of the effective cosmological constant. We also observe that the derivative
\beq
\label{tan4}
f'(R)=1-\frac{ab}{2\cosh^2\left(\frac{b\left(R-R_0\right)}{2}\right)}
\eeq
has a minimum when $R=R_0$, which reads
\beq
\label{tan5}
f'(R_0)=1-\frac{ab}{2}\, .
\eeq
In order to avoid antigravity one needs to require
\beq
\label{tan6}
0<f'(R)<f'(R_0)= 1-\frac{ab}{2}\, .
\eeq
The model given by Eq.~(\ref{tan1}) is able to describe late acceleration,
since, in general, de Sitter critical points exist. They are the solutions
of (\ref{B}). Then, we set
\beq
K(R)=2f(R)-Rf'(R)\,,
\label{RKR}
\eeq
and study (numerically) the zeros of this transcendental function
(see Fig.~\ref{P:tanh1}). We see that for suitable choices of the
parameters there are one or two de Sitter critical points.

\begin{figure}[ht]
\includegraphics[width=0.4\textwidth]{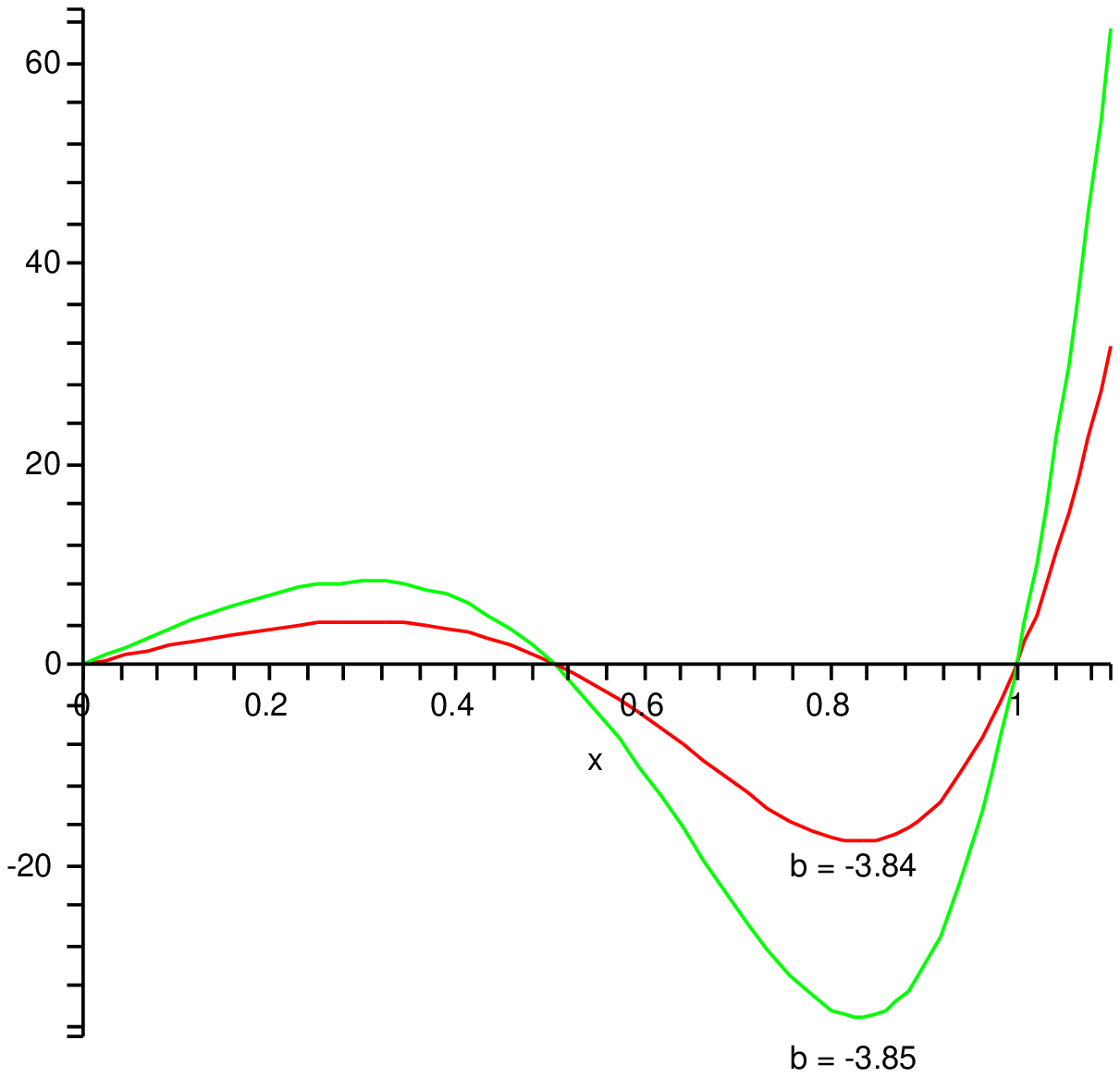}
\includegraphics[width=0.4\textwidth]{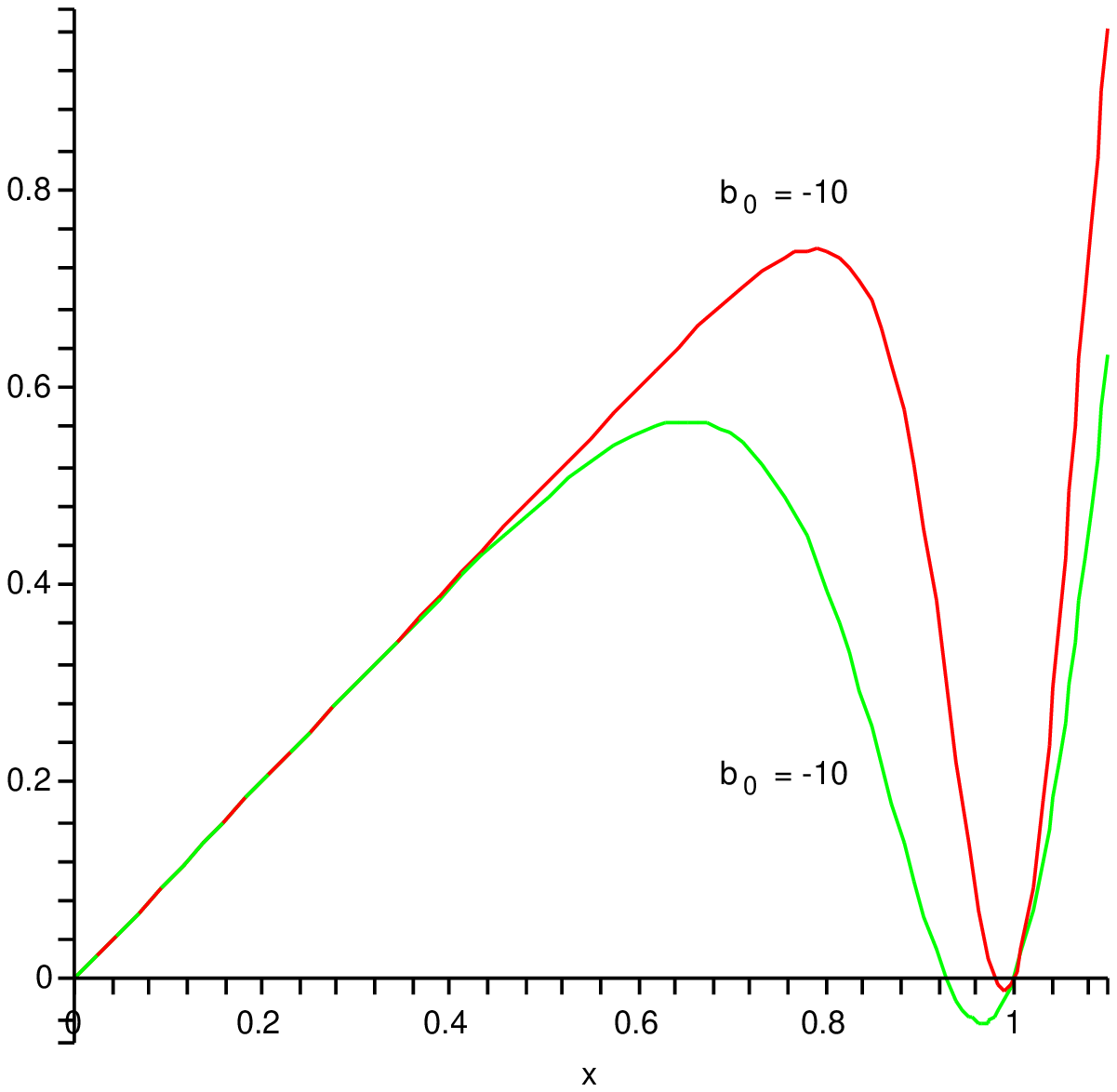}
 \caption{$K(R)$, with $R_0$ as a fixed de Sitter solution,
for different choices of the parameter $b_0$:
$b_0=-3.84$, $b_0=-3.85$ (picture on the left);
$b_0=-10$, $b_0=-20$ (picture on the right).}
 \label{P:tanh2}
 \end{figure}

Now, we impose $R_0$ to be a de Sitter critical point for the model in Eq.~(\ref{tan1}).
This means that (\ref{RKR}) has to be satisfied for $R=R_0$ and such a
condition fixes one parameter, say $a$. Then, we get
\beq
a=\frac{2R_0}{bR_0-4\tanh(bR_0/2)}\,,
\eeq
and introducing, for convenience, the dimensionless variables
\beq
x=\frac{R}{R_0}\,,\qquad\qquad b_0=R_0b\,,
\eeq
we obtain
\beq
f'(R_0)=1-\frac{b_0}{b_0-4\tanh(b_0/2)}\,.
\qquad\qquad
\Lambda_{\rm eff}=\frac{1+\tanh(b_0/2)}{b_0-4\tanh(b_0/2)}\,,
\eeq
\beq
K(R)=-\frac{(1-x)\tanh(b_0/2)+4\tanh(b_0(x-1)/2)+b_0x\tanh^2(b_0(x-1)/2)}
 {b_0-4\tanh(b_0x/2)}\,.
\eeq
In order to have $\Lambda_{\rm eff}>0$ and, at the same time, to avoid antigravity,
the parameter $b_0$ has to be negative and in the range $b_0<4\tanh(b_0/2)\sim-3.83$.
By varying $b_0$ in that range, the value of $\Lambda_{\rm eff}$ can
acquire any desired value. Moreover the model has always two de Sitter critical
points with constant curvatures $R_1$ and $R_2$. One can see that $R_2=R_0$, while $R_0/2<R_1<R_0$.
In Fig.~(\ref{P:tanh2}) we have plotted $K(R)$ for some values of $b_0$.
Thus, our study shows the existence of multiply-de Sitter universes in more
complicated, tangential models. One can prove that other viable $F(R)$
gravities introduced in ref.~\cite{Cognola:2007zu} naturally lead to
de Sitter solutions too, as demonstrated in Ref.~\cite{maria}.

\section{Black hole solutions and related thermodynamical quantities}

In this section we will discuss spherically symmetric exact solutions for the modified gravities
above. We mainly concentrate on multiply Black Hole solutions and on their related entropies and
free energies.

We have seen in the introduction that, requiring constant curvature
solutions, one may consider the simplified form of the equations of motion given in (\ref{E}).
As a result, $f(R)$ modified gravity models admit the following
class of general static neutral black hole
solutions---in four dimensions and with a non vanishing cosmological
constant. In order to describe them, recall the metric
\be
ds^2=-V(r)dt^2+\frac{dr^2}{V(r)}+r^2d\Sigma_k^2\,,\qquad V(r)>0\,,
\label{ds}
\ee
where $k=0,\pm 1$ and the possible horizon manifolds are $\Sigma_1=S^2$, the
two dimensional sphere, $\Sigma_0=T^2$, the two dimensional torus,
and $\Sigma_{-1}=H^2/\Gamma$, the two dimensional compact Riemann
surface.
The scalar curvature for the {\it ansatz} (\ref{ds}) reads
\be
R=-\frac{1}{r^2}[ r^2V''+4rV'+2V-2k ] \ee
and this means that every constant curvature solution
with $R=R_0$ has to satisfy the equation
\be
r^2V''+4rV'+2V-2k=-r^2 R_0\,.
\ee
The general solution of this differential equation depends on two
integration constants, $b$ and $c$, and reads
\be V(r)=\frac{b}{r^2}+k-\frac{c}{r}-\frac{R_0}{12}\,r^2\,.
\ee
The Ricci tensor and the scalar curvature have to satisfy the equations
\be
R_{\mu\nu}=\frac{R_0}{4}\,g_{\mu\nu}\,,\quad R=R_0\,,
\label{E1}
\ee
where $R_0$ is implicitly given by (\ref{B}), that is
\be
R_0=\frac{2f(R_0)}{f'(R_0)}\,,
\ee
Now, it is easy to verify that Eqs.~(\ref{E1}) are satisfied only if
$b=0$, while $c$ is an arbitrary parameter,
which is usually assumed to be non-negative and related to
the mass $M$ of the black hole by $c=2MG$. The special case $c=0$ is also admissible.
Then, we have
\be
V(r)=k-\frac{c}{r}-\frac{R_0}{12}r^2\,,\quad c\geq0\,,\quad V(r)>0\,.
\label{Vr}
\ee
In principle there are physical solutions of the latter equation for $k=0$
and $k=-1$, which give rise, respectively, to a torus topology and a
hyperbolic topology for the horizon manifold (the so called
topological black holes \cite{Vanzo:1997gw}).
Here we are mainly interested in the usual spherical symmetric horizons and so
we only consider in detail the case $k=1$.

As it is well known, in the special case $k=1,c=0$, $V(r)$ in (\ref{Vr})
is always positive when $R_0<0$, and this corresponds to the Anti de Sitter (AdS)
solution. On the contrary, when $R_0>0$, $V(r)$ is positive,
for $r<2\sqrt{3/R_0}$, and this corresponds to the de Sitter solution.
If $k=1$ and $c=2MG>0$, one has black hole solutions but only if
\be
c^2R_0-\frac{16}{9}\leq0\:\:\Longrightarrow\:\:\alpha\equiv\frac32\,MG\sqrt{R_0}\leq1\,.
\label{bkc}
\ee
We see that (\ref{bkc}) is always satisfied if $R_0<0$. This corresponds
to the Schwarzschild-Anti-de Sitter (SAdS) black hole. In this case $r>r_H$,
$r_H$ being the positive root of $V(r)=0$ (horizon radius).
If $R_0>0$, there are solutions only if $\alpha<1$. In this case
$r_H\leq r\leq r_C$, $r_H$ and $r_C$ being the positive roots of $V(r)=0$
(horizon and cosmological radius respectively).
The extremal case $\alpha=1$ is also admissible (Nariai solution),
but the thermodynamics of such a black hole have to be discussed
separately \cite{Bousso:1997wi}.

At this point, we provide a brief discussion regarding the thermodynamical
properties of the above black hole solutions. If one make use of the Noether
charge method for evaluating the entropy associated with the black hole
solutions with constant curvature \cite{Wald:1993nt} in modified $f(R)$
gravity models, one has \cite{B}
\be S=\frac{A_H}{4G} f'(R_H)\,. \label{ve}
\ee
where the factor $f(R)$ has to be evaluated on the horizon, with area $A_H$.
As a consequence, one obtains a modification of the ``Area Law''. Several examples
have been discussed in \cite{B}. In the case of a constant curvature
solution one has simply $R_H=R_0$, $R_0$ being
the solution of Eq.~(\ref{B}), which has been investigated in previous
Sections. With regard to this, we have found sufficient conditions to have
two de Sitter solutions. Thus, we may investigate their thermodynamical
behavior evaluating the associated free energy.

The free energy $\cal F$ is related to
the canonical partition function $Z$ by
\be
{\cal F}=-\frac{\log Z}{\beta}\,.
\ee
On the other hand, a semiclassical approximation gives
\be
Z\simeq e^{-I_E}\,,
\ee
where $I_E$ is the Euclidean classical action associated with the de Sitter
solution. A direct calculation leads to
\be
I_E=-\frac{24\pi f(R_0)}{G R_0^2}
\ee
and from (\ref{B}) and (\ref{ve}) it directly follows that
\be
I_E=-S_H\Longrightarrow {\cal F}=-\frac{S_H}{\beta_H}\,,
\label{fe}
\ee
$S_H$ and $\beta_H$ being, respectively, the entropy and the inverse
temperature of the black hole.

For a generic Schwarzschild-de Sitter (SdS) solution, Eq.~(\ref{fe}) reads
\be
{\cal F}=-\frac{2\pi T_H r_H^2 f(R_0)}{GR_0}\,.
\label{fe1}
\ee
The temperature $T_H$ is related to the horizon radius $r_H$ by
\be
T_H=\frac{1}{\beta_H}=\frac{|V'(r_H)|}{4\pi}=\frac{1}{4\pi}
\left|\frac{2MG}{r_H^2}-\frac{r_HR_0}{6}\right|
\ee
$r_H$ being a positive solution of the algebraic equation
\be
r_H^3-\frac{12}{R_0}r_H+\frac{24MG}{R_0}=0\,.
\label{ho}
\ee
As a result, we finally have
\be
{\cal F}=-\left|r_H-3MG\right|\,\frac{f(R_0)}{GR_0}\,.
\label{fe2}
\ee

In the pure de Sitter case $M=0$, $r_H=2\sqrt{3/R_0}$, and so one has
\be
{\cal F}_{dS}=-2\sqrt{3}\,\frac{f(R_0)}{GR_0^{3/2}}\,.
\label{FdS}
\ee
For SdS one needs to consider separately the two admissible cases
$\alpha=1$ and $\alpha<1$. Here we discuss the
second one only, that is, the proper SdS solution where
$\alpha=(3/2)MG\sqrt{R_0}<1$. In this case, Eq.~(\ref{ho}) has
one negative root and two distinct positive roots,
$r_H$, the event horizon and $r_C$, the cosmological horizon, with
$r_H<r_C$. The positive roots can be written in the form
\be
r_C=\frac{4\gamma_C}{\sqrt{R_0}}\,,\quad\quad \frac12<\gamma_C<1\,,
\ee
\be
r_H=\frac{4\gamma_H}{\sqrt{R_0}}\,, \quad\quad 0<\gamma_H<\frac12\,,
\ee
where
\be
\gamma_C=\frac12\,\left[\frac1{\left(\sqrt{\alpha^2-1}-\alpha\right)^{1/3}}
+\left(\sqrt{\alpha^2-1}-\alpha\right)^{1/3}
\right]
\ee
and
\be
\gamma_H=\frac12\,\left(\sqrt3\,\sqrt{1-\gamma_C^2}-\gamma_C\right)\,.
\ee
In this way, we get
\be
{\cal F}_C=\frac{|2\gamma_C-\alpha|}{\sqrt3}\,{\cal F}_{dS}\,,
\label{FC}
\ee
\be
{\cal F}_H=\frac{|2\gamma_H-\alpha|}{\sqrt3}\,{\cal F}_{dS}\,,
\label{FH}
\ee
${\cal F}_{dS}$ being the expression (\ref{FdS}).
It has to be noted that, in the latter equations,
the factors in front of ${\cal F}_{dS}$ are always smaller than 1
and thus the energies ${\cal F}_C$ and ${\cal F}_H$ are always smaller than
${\cal F}_{dS}$, independently of the value of the mass.
Moreover, the factor $|2\gamma_C-\alpha|/\sqrt3$ is a monotone decreasing function
of $\alpha$, which is equal to 1 for $\alpha=0$ and to 0 for $\alpha=1$,
while the factor $|2\gamma_H-\alpha|/\sqrt3$ is exactly equal to 0 for
$\alpha=0,1$ and quite small otherwise.
Its maximum value, a little bit more than $1/10$,
is reached for $\alpha\sim0.8$.
This means that the minimum value for the
free energy ${\cal F}_H$ of a SdS black hole is obtained for
$M\sim 0.53/(G\sqrt{R_0})$.

\begin{figure}[ht]
\includegraphics[width=0.4\textwidth]{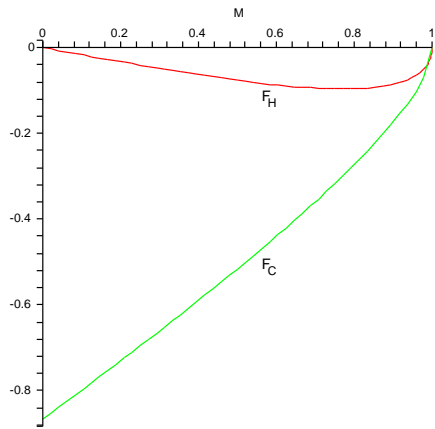}
 \caption{${\cal F}_C$ and ${\cal F}_H$
 SdS free energies for the $\Lambda$CDM model.}
 \label{P:LaCDM}\end{figure}

Now, we shall study some models and explicitly compute the corresponding free energy.
First of all, as a trivial example, we consider the $\Lambda CDM$ model
described by $f(R)=R-2\Lambda$. This has a stable de Sitter solution, with
$R_0=4\Lambda$. Then, we immediately have
\be
{\cal F}_{dS}=-\frac{\sqrt3}{2G\sqrt\Lambda}.
\ee
The corresponding SdS cosmological and black-hole free energies
are plotted in Fig.~\ref{P:LaCDM}, as a function of the mass $M$
(in energy units $G\sqrt\Lambda$).

As a second example we consider the viable model in (\ref{1.5}), with
$k=4$, $c_1=1.12\cdot10^{-3}$, $c_2=6.21\cdot10^{-5}$, $c_3=0$, as discussed in Sect.~II.
With this choice of parameters, Eq.~(\ref{B}) has three real solutions for
$R_0=0$ (Minkowski), $R_1\simeq15.33\mu^2$ and $R_2\simeq34.92\mu^2$.
The last one ($R_2$) is stable, while the other two are unstable. For the
free energies of the two de Sitter solutions we get
\be
{\cal F}^{(1)}_{dS}=-2\sqrt3\,\frac{f(R_1)}{R_1^{3/2}}
 \simeq-\frac{0.078}{\mu G}\,,\quad\quad
{\cal F}^{(2)}_{dS}=-2\sqrt3\,\frac{f(R_2)}{R_2^{3/2}}
 \simeq-\frac{0.286}{\mu G}\,.
\label{minimoDS}
\ee
We see that
\be
{\cal F}^{(2)}_{dS}<{\cal F}^{(1)}_{dS}\,,\qquad\qquad
\frac{{\cal F}^{(2)}_{dS}}{{\cal F}^{(1)}_{dS}}\sim 3.636\,.
\ee
independently of the parameter $\mu$ and so
the stable de Sitter solution with curvature $R_2$
is favorite, from the energetic point of view,
with respect to the one with smaller curvature $R_1$.

\begin{figure}[ht]
\includegraphics[width=0.4\textwidth]{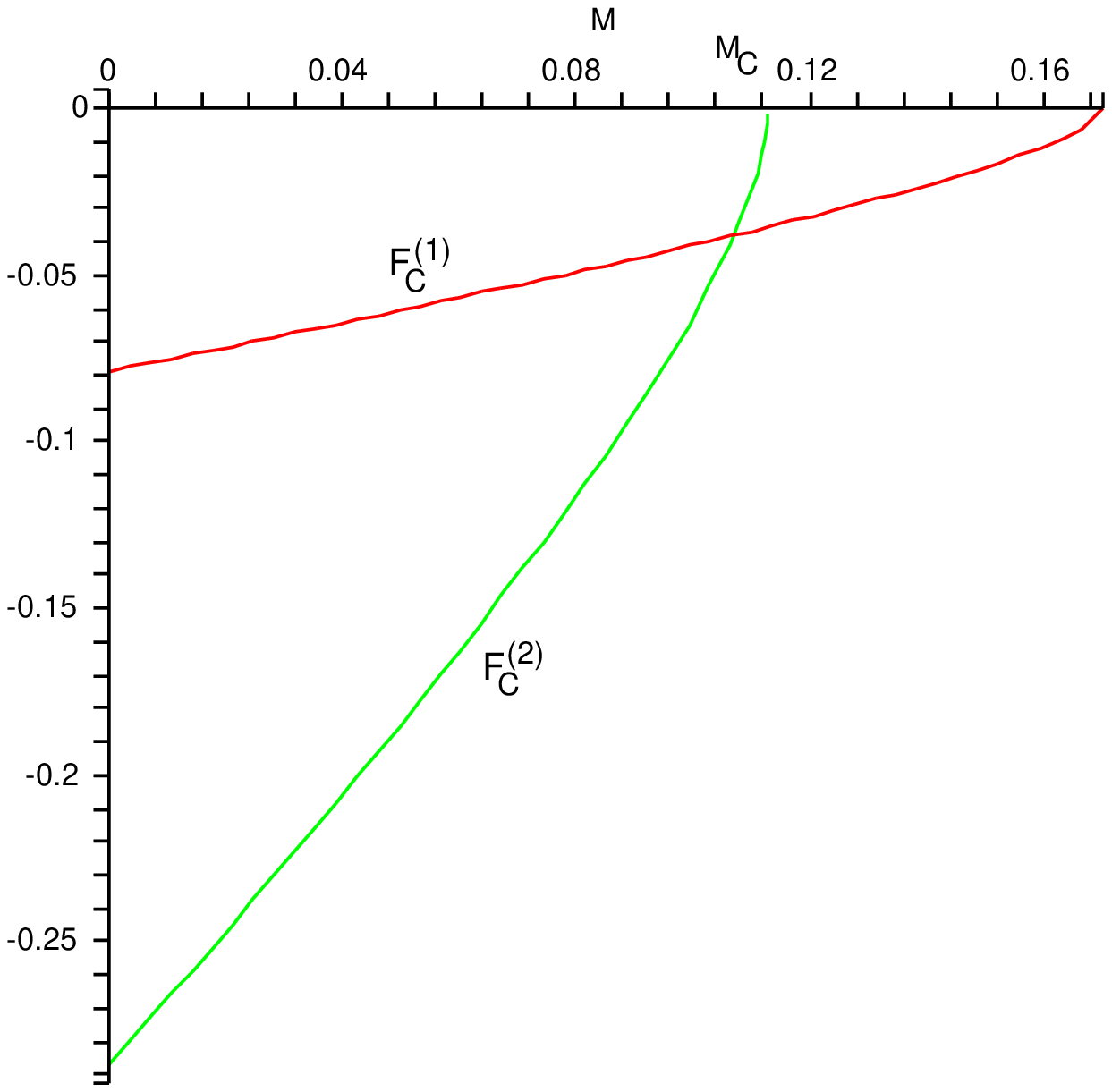}
\includegraphics[width=0.4\textwidth]{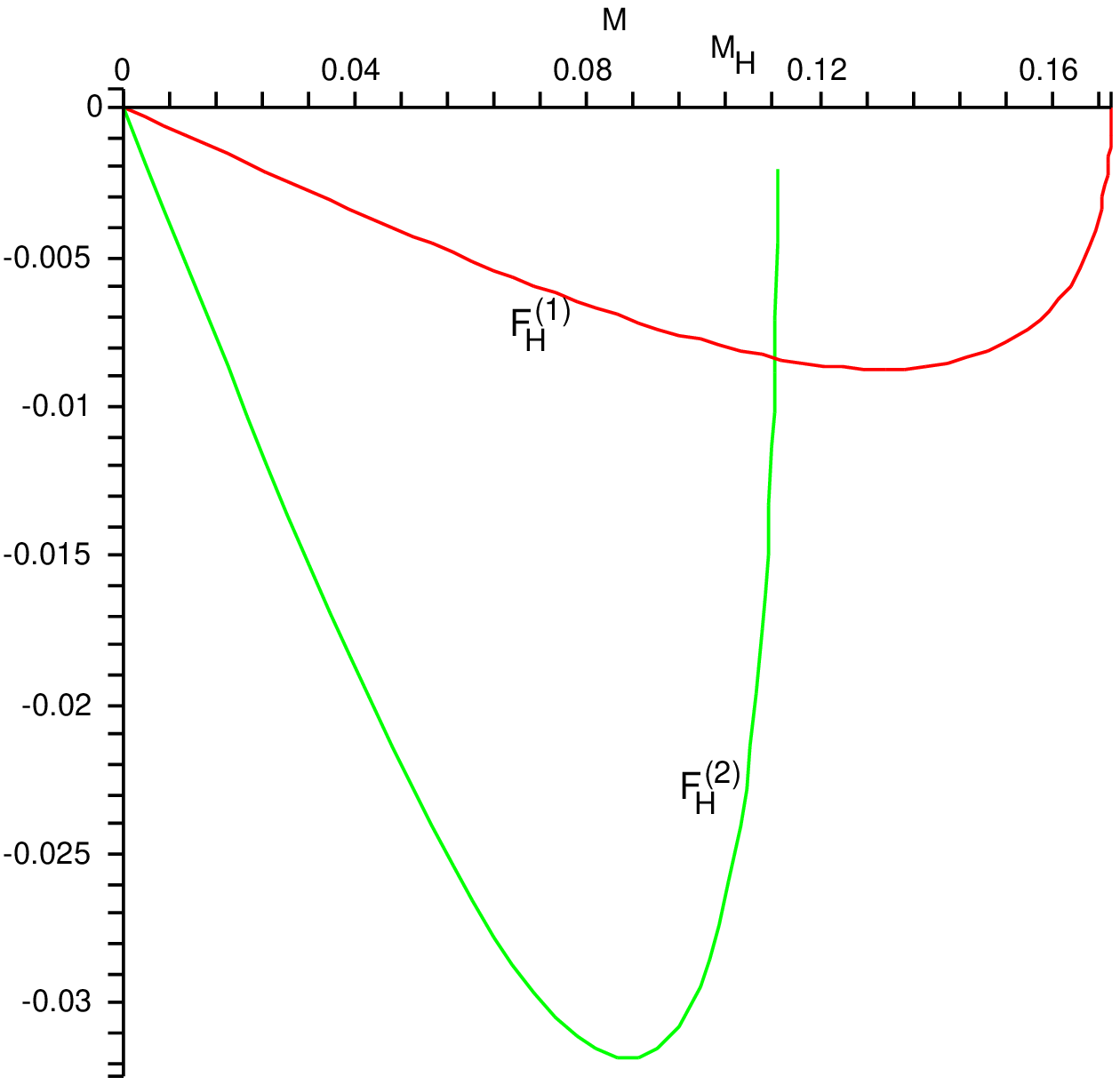}
 \caption{SdS free energies for the two de Sitter critical points
 of the model (\ref{1.5}):
 cosmological free energies on the left and
 black-hole free energies on the right.}
\label{P:FcFh}\end{figure}

The SdS free energies are plotted in Fig.~\ref{P:FcFh}, as functions
of the mass $M$ (in energy units $1/\mu G)$, for
$0\leq M<2/(3G\sqrt{R_1})$. However, it has to be stressed that,
for the SdS corresponding to the de Sitter solution
with curvature $R_2$, the mass needs to be restricted to
$0<M<2/(3G\sqrt{R_2})<2/(3G\sqrt{R_1})$ and thus,
for a fixed mass $M$, we can compare the corresponding free energies in this
range only. For both the cosmological and black-hole horizons
there are critical values of the black hole mass $M$, say $M_C<M_H$,
for which ${\cal F}^{(1)}_{C,H}={\cal F}^{(2)}_{C,H}$.
Then, we can distinguish three different regions:\\
(a) $M<M_C$: in this first case
 ${\cal F}^{(2)}_{C,H}<{\cal F}^{(1)}_{C,H}$;\\
(b) $M_C<M<M_H$: in this second case
 ${\cal F}^{(2)}_{C}<{\cal F}^{(1)}_{C}$,
 but ${\cal F}^{(2)}_{H}>{\cal F}^{(1)}_{H}$;\\
(c) finally, for $2/(3G\sqrt{R_1})>M>M_H$,
 we have ${\cal F}^{(2)}_{C,H}>{\cal F}^{(1)}_{C,H}$.

As we have seen from general considerations,
the free energies corresponding to the
cosmological horizons are monotonous functions of the mass,
while the ones corresponding to the black-hole horizons
are convex functions, which reach the minimum at $M\sim0.53/(G\sqrt{R_0})$.
For our two cases, $R_1$ and $R_2$, we get
\be
M_1\sim\frac{0.53}{G\sqrt{R_1}}\sim\frac{0.135}{\mu G}\,,
 \quad\quad
 {\cal F}^{(1)}_{H,min}\sim-\frac{0.008}{\mu G}\,,
\ee
\be
M_2\sim\frac{0.53}{G\sqrt{R_2}}\sim\frac{0.089}{\mu G}\,,
 \quad\quad
 {\cal F}^{(2)}_{H,min}\sim-\frac{0.032}{\mu G}\,.
\label{minimum} \ee Then we see that, for the model in (\ref{1.5}) we are dealing with, the configuration with the minimum free energy corresponds to SdS with mass
$M=M_2\sim0.089/(\mu G)$ and
curvature $R=R_2\sim/34.92\mu^2$. The associated free energy is given in (\ref{minimum}).
We also observe that this is smaller than the free energy of the pure de Sitter configuration as it is clear from (\ref{minimoDS})

Thus, it is demonstrated here that multiple de Sitter solutions can appear also
under the form of SdS solutions.
In other words, the number of multiply solutions becomes significantly bigger.
With the appearance of both dS and SdS universe solutions one can suggest
various scenarios for the universe evolution.
For instance, we can conjecture that a (pre-)inflationary universe is
described by some SdS spacetime. As time proceeds, this universe decays and enters into the well known radiation/matter dominance phase. In its further
evolution, the universe transits to the dS (or almost dS) era, by stability
and least-energy principle considerations. The future universe may again appear as an
SdS spacetime.

\section{Discussion}.

In summary, we have investigated in this paper several viable models of 
modified gravity which satisfy both the constraints of local as 
well as cosmological tests. By means of a numerical study, it is demonstrated that
some versions of highly non-linear models exhibit multiply de Sitter
universe solutions which often appear in pairs, being one of them stable
and the other unstable. The numerical evolution of the effective
equation of state parameter is presented too.  As a result,
these models can be considered as natural candidates for the unification of early-time
inflation with late-time acceleration through dS critical points. Moreover,
based on the de Sitter solutions, multiply SdS solutions can also be constructed.
Further, we have investigated the thermodynamic properties of these SdS universes: their corresponding entropies and free energies have been calculated and compared. SdS universe might also appear at the (pre-)inflationary stage.

Owing to the highly non-linear structure of the theories under discussion,
the dS universes had to be constructed with numerical tools mainly. Moreover, in order to simplify the problem, at this first stage of the investigation we did not to consider matter contributions. It is clear that, in the next step,
we must necessarily include matter and reconsider the problem in its presence.
This has the potential to lead to a sufficiently realistic quantitative description of the universe expansion history, in which modified gravity would be responsible for both
acceleration stages: the inflation epoch and the dark energy one. This quantitative
analysis will be presented elsewhere.

\noindent {\bf Acknowledgements.} This paper is an outcome of the collaboration program INFN (Italy)
and DGICYT (Spain). It has been also supported in part by MEC (Spain), projects FIS2006-02842 and
PIE2007-50I023, by AGAUR (Gene\-ra\-litat de Ca\-ta\-lu\-nya), contract 2005SGR-00790 and grant
2008BE1-00180, and by RFBR, grant 06-01-00609 (Russia). The work of P.T. was partially supported by
RFBR, grant 08-02-00923, and with the scientific school grant 4899.2008.2 of the Russian Ministry of
Science and Technology. PT thanks Alexey Toporensky for some useful discussions. The research of EE
is partly based on work done while on leave at the Department of Physics and Astronomy, Dartmouth
College, 6127 Wilder Laboratory, Hanover, NH 03755, USA.

\end{document}